\documentstyle[aps,12pt]{revtex}

\begin{document}
\title{The 3$d$-electron states in FeBr$_{2}$}
\author{Z. Ropka}
\address{Center for Solid State Physics, S$^{nt}$ Filip 5,31-150Krakow,Poland%
}
\author{R. J. Radwanski}
\address{Center for Solid State Physics, S$^{nt}$ Filip 5,31-150Krakow,Poland%
\\
Institute of Physics, Pedagogical University, 30-084 Krakow, Poland\\
email: sfradwan@cyf-kr.edu.pl, http://www.css-physics.edu.pl}
\maketitle

\begin{abstract}
The fundamental controversy about the electronic structure for 3$d$-electron
states in FeBr$_{2}$ is discussed. We advocate for the localized electron
atomic-like many-electron crystal-field approach that yields the discrete
energy spectrum, in the scale of 1 meV, associated with the atomic-like
states of the Fe$^{2+}$ ions in contrary to the (semi-)continuous energy
spectrum yielded by band theories. In our approach the six {\it d} electrons
of the Fe$^{2+}$ ion form the highly-correlated atomic-like electron system 3%
$d^{6}$ described by two Hund's rules quantum numbers $S$=2 and $L$=2 with
taking into account the spin-orbit coupling. The superiority of our model
relies in the fact that it explains consistently properties of FeBr$_{2}$,
the insulating and the magnetic ground state as well as thermodynamics and
Raman low-energy spectra, using well-established physical concepts.

PACS\ No: 75.50.Ee : 71.15.Mb

\end{abstract}

\date{(Receipt: 18 June 2002)}

There is a fundamental controversy in the description of 3$d$ states in FeBr$%
_{2}.$ This controversy is related to origin of the magnetism and of the
electronic structure of 3$d$-atom containing compounds. The understanding of
the magnetism and of the electronic structure is very closely related to a
general problem ''how to treat $d$ electrons in 3$d$-atom containing
compounds''. The fundamental controversy ''how to treat $d$ electrons''
starts already at the beginning - should they be treated as localized or
itinerant (in the scientific physics sense, i.e. the localized or the
itinerant picture should be taken as the start). Directly related to this
problem is the structure of the available states: do they form the
continuous energy spectrum like it is in the band picture or do they form
the discrete energy spectrum typical for the localized states. Recently, at
the beginning of 2002, Youn, Sahu and Kim (YSK) have presented an
all-electron fully relativistic density-functional calculations, an
extension of the density-functional approximation (LDA), for the electronic
structure and the resulting magnetic moment, including the orbital moment %
\cite{1}. This paper followed a paper of \ Ropka, Michalski and Radwanski
(RMR), from May 2001, who have presented, within the quasi-atomic approach,
the electronic structure and the resulting properties like spin and orbital
contributions to the magnetic moment of the iron atom and thermodynamics %
\cite{2}. YSK claim, contradicting the earlier RMR result for the spin-orbit
coupling origin of the orbital moment, that the orbital moment is rather due
to the strong on-site correlations of the 3$d$ orbitals.

The aim of this paper is to show up that both approaches largely concurs as
far as the origin of the orbital moment is concerned. However, there is a
fundamental difference between these two approaches as far as the
description of the electronic structure and the physical nature of the 3$d$
states is discussed. According to us, this fundamental difference in the
electronic structure can be experimentally revealed and can be used as the
experimental verification of different theoretical approaches.

The electronic structure for the 3$d$ electrons in the YSK approach is
related to the band description. Youn et al. have extended the standard
local density approximation (LDA), in fact the LSDA, calculations by
introducing a large intraorbital correlation energy U of the $d$ electrons.
Such the calculations are known as L(S)DA+U calculations \cite{3,4,5,6,7}.
Their L(S)DA+U results for the $d$-electron states are schematically shown
in Fig. 1. The energy spectrum for the 3$d$ states contains continuous,
though quite narrow, bands centered at -9 eV, -6.5 eV (built up from 5
spin-up states) and at -0.2, +1.6 and +2.2 eV for the spin-down states.
Thus, the occupied 3$d$ states spread over $\sim $10 eV. A gap, needed for
the insulating state, within this band picture is obtained by the
correlation energy U \cite{1}.

The general shape of the bands presented in Fig. 1 can be somehow understood
knowing the localized states within the single-electron crystal field. The
continuous energy spectrum looks like the smooth convolution on the
available, largely spread, localized 3$d$ one-electron orbitals, $t_{2g}$
and higher $e_{g}$ orbital. According to this picture 6 electrons (spins) of
the Fe$^{2+}$ ion are put subsequently one by one on the one-electron states
formed for one $d$ electron in the octahedral crystal field (CEF): three
spins are put to the spin-up $t_{2g}$ orbitals, other two occupies spin-up $%
e_{g}$ orbitals, and the last electron has to occupy one spin-down t$_{2g}$
orbital. According to Refs \cite{1} such the electron/spin arrangement
satisfies the first Hund's rule.

Ropka et al.'s description, Ref. \cite{2}, of the electronic structure, Fig.
2, is related to the localized picture, but with taking into account {\bf two%
} Hund's rules and the intra-atomic spin-orbit coupling. In this atomic-like
many-electron CEF approach the 6 {\it d} electrons in the open 3{\it d}
shell form the highly-correlated intra-atomic 3$d^{6}$ electron system. The
strong correlations among the 3$d$ electrons are accounted for, in a
zero-order approximation, by two Hund's rules. The two Hund's rules yield
for the $d^{6}$ system the ground term $^{5}D$ with $S$=2 and $L$=2. Its
25-fold degeneracy is removed by the crystal field and spin-orbit
interactions. In the octahedral bromium-anion surrounding of the Fe$^{2+}$
ion, realized in the hexagonal structure of FeBr$_{2}$, the orbital triplet $%
^{5}T_{2g}$ is lower than the orbital doublet $^{5}E_{g}$, being separated
by $\Delta $ of about 2 eV. The $T_{2g}$ ground subterm is confirmed by our 
{\it ab initio} calculations of the hexadecapolar potential (the $A_{4}$ CEF
coefficient) at the Fe site in the bromium-anion octahedron. The $A_{4}$ CEF
coefficient in the center of the anion octahedron is positive. The $A_{4}$
CEF coefficient together with the Stevens factor $\beta $ for the 3$d^{6}$
system, of +2/63 (\cite{8}, table 19 on page 873), yields the positive value
of the octahedral CEF parameter $B_{4}$ and the $^{5}T_{2g}$ ground subterm.
The positive sign for $B_{4}$ in the anion octahedron has been derived
already a pretty long time ago by Abragam and Bleaney (\cite{8}, see page
374). However, this result has been completely forgotten in the current
theoretical solid-state physics.

In general, we take the VSK results as the confirmation of our result about
the substantial orbital moment. We fully agree that the orbital moment is
mostly due to the correlations of the 3$d$ orbitals, though the YSK paper
tries to contradict both our results about the origin of the large orbital
magnetic moment pointing out that Ropka et al. have speculated (no, no, we
have calculated it, and we can prove it!!!) that the orbital moment is due
to the spin-orbit coupling. According to us, our statement about the
spin-orbit coupling origin of the orbital moment does not exclude the
correlation effects. In contrast, our many-electron CEF approach considers $d
$ electrons in the very strong correlation limit. It was, somehow, correctly
recognized by VSK that in the quasi-atomic model the {\it d-d} on-site
correlations are included inherently (by two Hund's rules), despite of their
lack in the explicit form.

The obtained value of the ordered magnetic moment at zero temperature
amounts to 4.26 $\mu _{B}$. It is built up from the spin moment of +3.48 $%
\mu _{B}$ and from the orbital moment of +0.78 $\mu _{B}$ \cite{2}. We point
out that the same sign of the spin and orbital moments comes out directly
from our calculations and is related to the negative sign of the
intra-atomic spin-orbit coupling. The effective moment calculated from the $%
\chi ^{-1}$ {\it vs} T plot in the 300-400 K region amounts to 5.39 $\mu _{B}
$. It is only slightly larger value than the spin-only value of 4.90 $\mu
_{B}$ what is a surprise owing to the fact that the full quantum orbital
value $L$ (=2) is taken into calculations.

Our description of the electronic structure of FeBr$_{2}$ proceeds within
the intermediate crystal-field regime in contrary to the current-literature
view that prefers the strong crystal field regime in the description of 3$d$%
-atom compounds. According to our picture, the crystal field is\
intermediate in the sense that it is stronger than the spin-orbit coupling
but it is relatively weak because it does not break the intra-atomic
arrangement of electrons within the 3$d$ shell, as is in the one-electron
version of the crystal-field theory. It means, that we think that the
intra-atomic structure of a paramagnetic atom largely perseveres even when
this atom becomes the full part of a solid - on this basis we have developed
a Quantum Atomistic Solid-State Theory (QUASST) for compounds containing
open 3$d$-/4$f$-/5$f$-shell atoms \cite{9,10}. This weak crystal-field
approach, known within the rare-earth CEF\ community as the CEF approach,
has been often successfully applied to 3$d$-ion doped systems, when 3$d$
ions were introduced as impurities, in interpretation of, for instance,
electron-paramagnetic-resonance experiments \cite{8,11,12}. In Refs \cite%
{13,14,15,16} we have applied this approach to a 3$d$-ion system where\
Co/Fe/Ni ions are the full part of a solid forming LaCoO$_{3}$, FeO or NiO.
Of course, in our picture the crystal field is much stronger than the
spin-orbit coupling as is generally accepted in the 3$d$ magnetism. It is
worth noting that the CEF states in our approach are many-electron states of
the whole highly-correlated electron system 3$d^{6}$. Moreover, in QUASST
the magnetic state, below T$_{N}$ of 14 K, is clearly distinguished from the
paramagnetic state, see Fig. 4 of Ref.\cite{2}. It is not clear how will
look like the band electronic structure of Ref.\cite{1} in the paramagnetic
state, i.e. above T$_{N}$ of 14 K, because it was not presented and not
discussed in Ref.\cite{1}.

We would like to point out that the QUASST approach should not be considered
as the treatment of an isolated ion - we consider the cation in the
octahedral crystal field. This octahedral crystal field is predominantly
associated with the anion octahedron FeBr$_{6}$. The FeBr$_{2}$ structure is
built up from the face sharing octahedra FeBr$_{6}$ - thus such the atomic
structure occurs at each cation due to the translational symmetry. The
strength of the crystal-field interactions is determined by the whole charge
surroundings, not only by the nearest bromium octahedron. It makes that the
CEF approach, in the version applied in QUASST, looks like a single-ion
approach but in fact it describes the coherent states of the whole crystal.
The parameters used in the discussion of properties of FeBr$_{2}$ in Ref. %
\cite{2}, are fully physical: the octahedral CEF parameter B$_{4}$ =+200 K,
the trigonal B$_{2}^{0}$ = -30 K and the spin-orbit coupling $\lambda $=
-150 K. They yield the overall effect of 2.06, 0.06, and 0.13 eV,
respectively.

The electronic structure can be experimentally verified. In QUASST the shown
states manifest in many physical properties like temperature dependence of
the heat capacity, of the paramagnetic susceptibility $\chi $, of the
magnetic-moment value. The energy separations should be detected by energy
spectroscopy experiments like Raman spectroscopy, for instance. In fact, we
take old, but rather unknown, results of the Raman spectroscopy \cite{17,18}
revealing a number of well-defined energy excitations as a confirmation of
our atomistic model. The observed 8 excitations up to 637 cm$^{-1}$ can be
fully reproduced by our calculations with the same crystal-field parameters B%
$_{4}$=+200 K and B$_{2}^{0}$=-30 K and a spin-orbit coupling $\lambda $ of
-161 K: 0 (exp. - 0); 22.8 cm$^{-1}$ (exp: - 23 cm$^{-1}$); 181 (192); 209
(244); 281 (296); 516 (517); 517 (517); 608 (584); 630 (630); 636 (637 cm$%
^{-1}$) \cite{19}. The fitted value for the spin-orbit coupling in FeBr$_{2}$
is only slightly larger than the assumed by us free-ion value of -150 K \cite%
{20}. So good description of new type of experimental results with the
parameters derived from analysis of the heat capacity and of the magnetic
moment we take as a strong argument for the high physical adequacy of
QUASST. \ 

In conclusion, we point out that the band L(S)DA description of FeBr$_{2}$,
completed by the U term, largely concurs to the localized many-electron CEF
description in revealing the very strong electron correlations determining
the orbital magnetic moment. These two approaches can, however, be easily
distinguished, as the proposed electronic structures are completely
different. Also the nature of the 3$d$ states is completely different. We
are in favour of the localized electron atomic-like many-electron
crystal-field approach that yields the discrete energy spectrum associated
with the atomic-like states of the Fe$^{2+}$ ions. The superiority of our
model relies in the fact that it consistently describes both
zero-temperature properties of FeBr$_{2}$ and thermodynamics and it makes
use of the well-defined physical concepts. Our atomic-like approach provides
in the very natural way the insulating ground state for FeBr$_{2}$
independently on the lattice distortions. The success of the LDA+U
calculations of Youn et al. is related to the orientation of LDA
calculations to the strong intra-atomic {\it d-d} correlation limit.
According to us, the LDA+U calculations have to largely reproduce {\bf two}
Hund's rules. We welcome the paper of Ref.\cite{1}- it bridges somehow LDA
calculations and CEF calculations. However, we would like to put attention
that our electronic structure is determined within the meV\ scale what is
about 5000 times smaller than the band width of Ref. \cite{1} of 0.5 eV.

Fig. 1. Description of the $d$ states in FeBr$_{2}$ within the band approach
obtained within the L(S)DA+U approach of Youn et al. There is a
semi-continuous energy spectrum of the $d$ states spread over 10 eV.

Fig. 2. The fine electronic structure of the highly-correlated 3$d^{6}$
electronic system of the Fe$^{2+}$ ion in FeBr$_{2}$. a) the 25-fold
degenerated $^{5}$D termgiven by Hund's rules: $S$=2 and $L$=2. b) the
effect of the octahedral crystal-field, c) the combined action of the
spin-orbit coupling and the octahedral crystal field: B$_{4}$=+200 K, $%
\lambda $= -150 K; d) the electronic structure of the Fe$^{2+}$ ion in FeBr$%
_{2}$; the trigonal-distortion parameter B$_{2}^{0}$= -30 K produces a
spin-like gap of 2.8 meV; the double degeneracy of the ground state is
removed in the magnetically-ordered state below T$_{N}$ of 14 K.

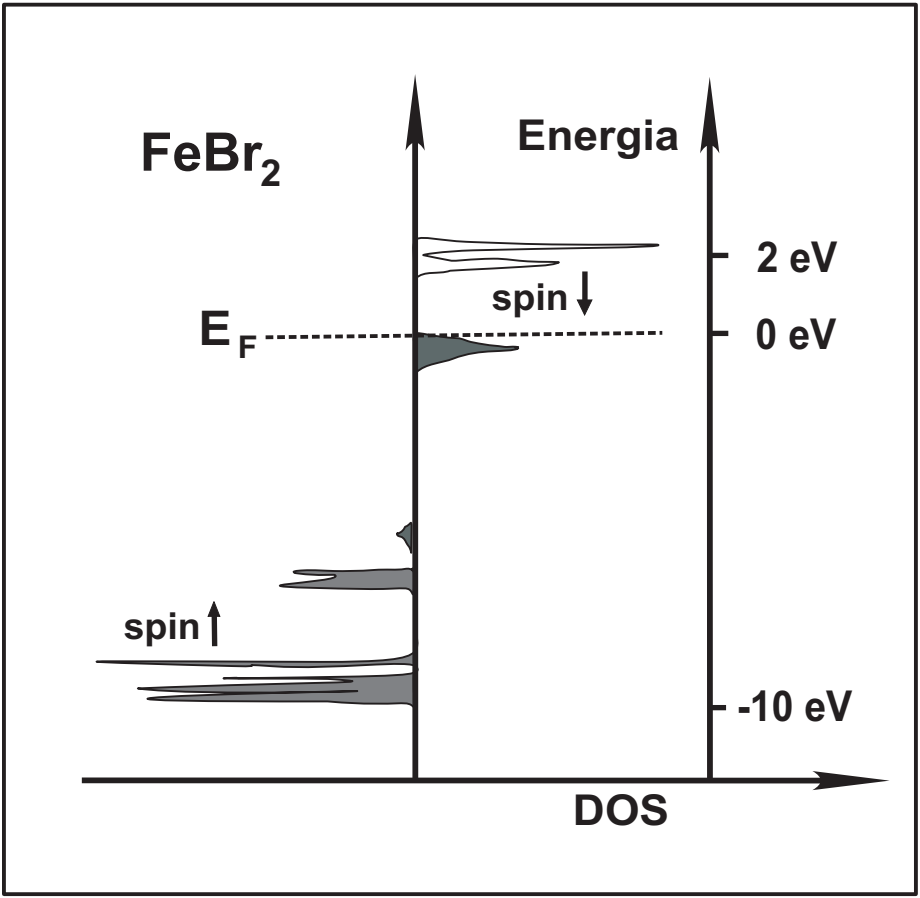
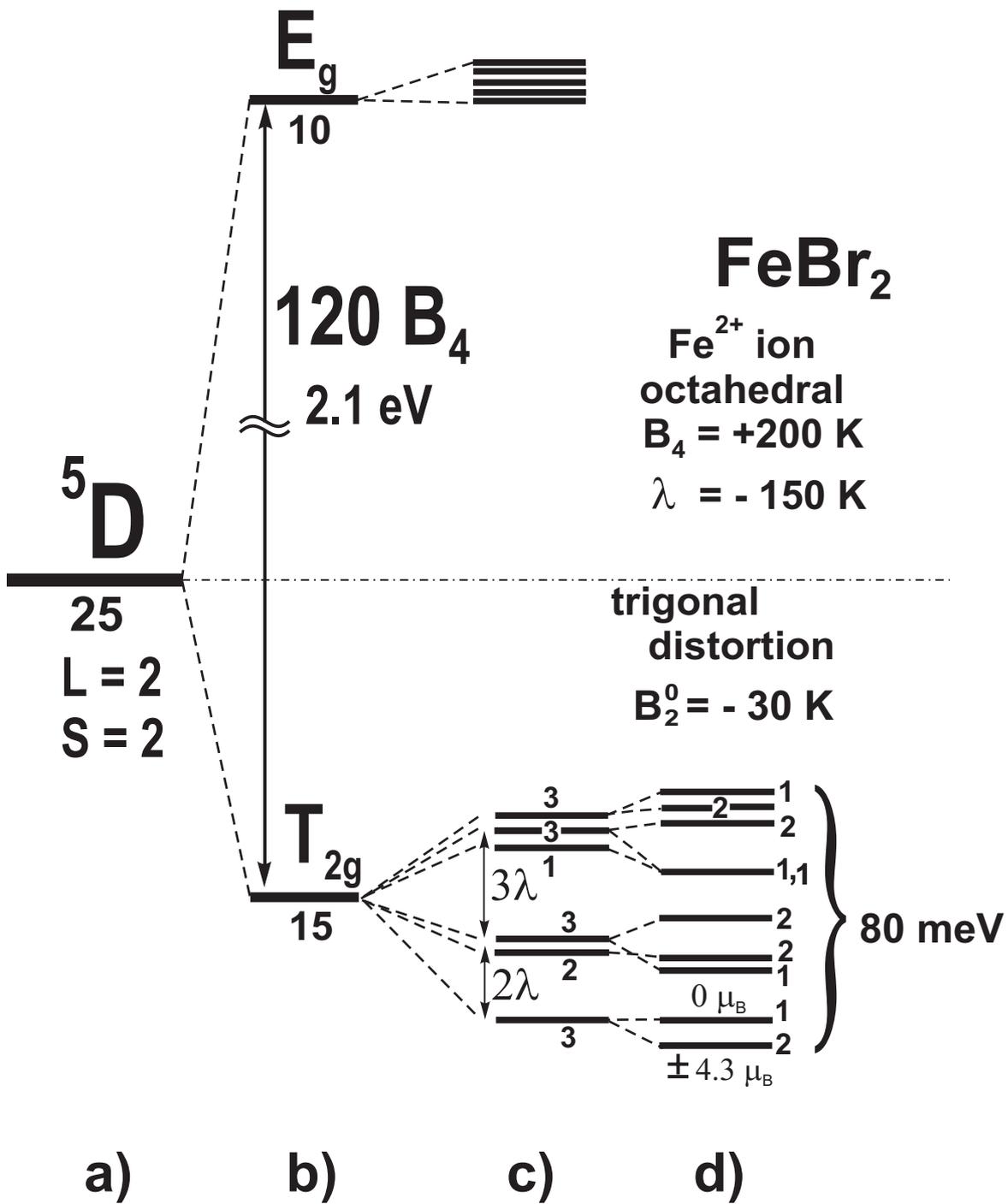
\end{document}